\documentclass[journal=jacsat,manuscript=article]{achemso}

\author{Stephan Schnez}
\email{schnez@phys.ethz.ch}
\phone{+41 (0)44 63 33792}
\author{Johannes G\"uttinger}
\author{Magdalena Huefner}
\affiliation[ETH Zurich]{Solid State Physics Laboratory, 8093 Zurich, Switzerland}
\author{Christoph Stampfer}
\affiliation[ETH Zurich]{Solid State Physics Laboratory, 8093 Zurich, Switzerland}
\alsoaffiliation[RWTH Aachen]{JARA-FIT and II. Institute of Physics, 52074 Aachen, Germany}
\author{Klaus Ensslin}
\author{Thomas Ihn}
\affiliation[ETH Zurich]{Solid State Physics Laboratory, 8093 Zurich, Switzerland}

\title{Imaging Localized States in Graphene Nanostructures}


\begin{document}
\begin{abstract}
  Probing techniques with spatial resolution have the potential to lead to a better understanding of the microscopic physical processes and to novel routes for manipulating nanostructures. We present scanning-gate images of a graphene quantum dot which is coupled to source and drain via two constrictions. We image and locate conductance resonances of the quantum dot in the Coulomb-blockade regime as well as resonances of localized states in the constrictions in real space.

\end{abstract}



Graphene has sparked intense research among theorists and experimentalists \cite{Geim07, Neto09} alike since its first successful fabrication in 2004 \cite{Novoselov04}. This is mainly due to graphene's extraordinary band structure, a linear relationship between energy and momentum without a band gap. The gapless band structure, however, prohibits confining charge carriers by using electrostatic gates. Hence, lateral confinement in graphene relies on etched structures and the appearance of a transport gap in graphene constrictions \cite{Ponomarenko08, Stampfer08, Liu09}. Nevertheless, already the first experiment on graphene nanoribbons by Han \emph{et al.} \cite{Han07} showed a discrepancy between the measured transport gap and a simple confinement-induced band gap. Theoretical models explain the observed gap by Coulomb blockade, edge scattering, and/or Anderson-type localization due to edge disorder \cite{Sols07, Mucciolo09, Evaldsson08, Schubert09}. On the experimental side, there is increasing evidence for Coulomb-blockade effects in nanoribbons \cite{Liu09, Todd08, Molitor09, Stampfer09, Gallagher09}. 

Transport through graphene quantum dots in the Coulomb blockade regime is typically modulated by resonances arising from the constrictions \cite{Stampfer08b}. However, for both, nanoribbons and quantum dots, the microscopic origin of the transport gap and the resonances in the constrictions needs to be understood in more detail. 

Hence, probing techniques which are capable of locally investigating properties of graphene nanostructures are essential. Earlier experiments of this kind on graphene include experiments with a scanning single-electron transistor \cite{Martin08, Martin09}, scanning-tunneling spectroscopy \cite{Zhang08, Zhang09, Li09}, and scanning-gate microscopy \cite{Berezovsky09, Connolly09}. All these experiments were performed on large area graphene sheets. Here, we present results of scanning-gate experiments on a single-layer graphene quantum dot which is coupled to source and drain leads via two constrictions. We observe ring-like resonances in the scanning-gate experiments centered at the quantum dot as well as in the constrictions. This enables us to map resonances in real space.


\begin{figure}
  \centering
  \includegraphics[width= 12 cm]{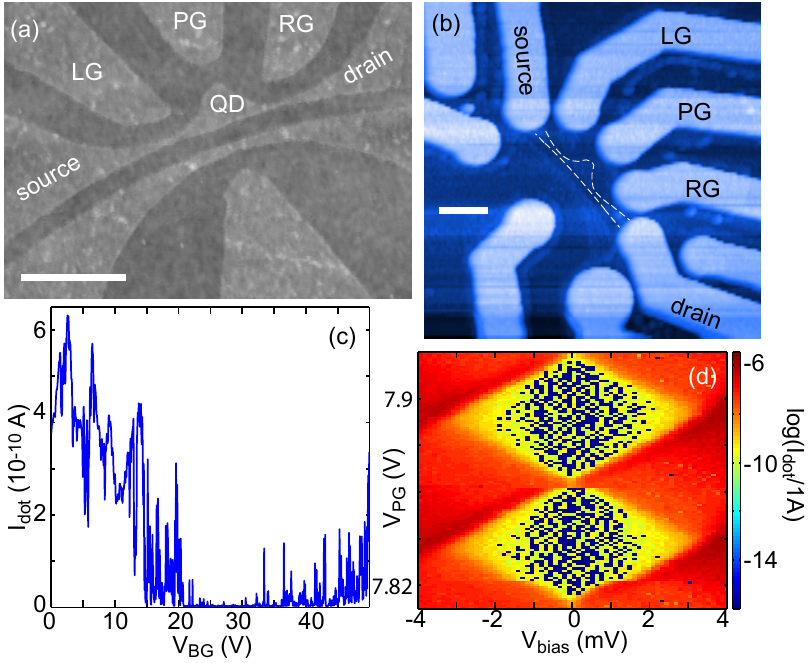}
  \caption{(a) Atomic force micrograph of the graphene sample after reactive ion etching obtained under ambient conditions. The quantum dot (QD) is connected to source and drain via two constrictions. The nearby nanoribbon can be used as a charge detector but it was not connected in the measurements presented here. (b) In-situ atomic force micrograph of the sample after cooldown at $T\approx 2.6\,\textrm{K}$. This image was taken after positioning the tip above the sample with our home-built AFM \cite{Gildemeister07}. Scale bars in (a) and (b) denote $500\,\textrm{nm}$. (c) Backgate trace taken at $T\approx 2.6\,\textrm{K}$ and $V_{\textrm{bias}}=500\,\mu\textrm{V}$. The charge neutrality point is shifted to $V_{\textrm{BG}}\approx 30\,\textrm{V}$. (d) Charge stability diagram of the quantum dot. The charging energy is found to be $\Delta E_C = 3.5\,\textrm{meV}$ at $V_{\textrm{BG}}=15\,\textrm{V}$ and $T=90\,\textrm{mK}$.\label{fig1}}
\end{figure}

Atomic-force micrographs of the sample after etching under ambient conditions (a) and of the completed device at $T\approx 2.6\,\textrm{K}$ (b) are shown in Fig.~1. If not stated otherwise, the temperature of all measurements shown in this paper is $T=2.6\,\textrm{K}$. Fabrication details are given in the supporting information \cite{supp_info}. We first show a backgate sweep in Fig.~1(c) with voltage $V_{\textrm{bias}}=500\,\mu\textrm{V}$ applied between source and drain. The current through the dot is suppressed in the transport gap ranging approximately from $15\,\textrm{V}$ to $45\,\textrm{V}$. The charge neutrality point is at $V_{\textrm{BG}}\approx 30\,\textrm{V}$, presumably because of charged impurities on or near the graphene surfaces. The charge stability diagram of the quantum dot in Fig.~1(d) was measured at the base temperature $T=90\,\textrm{mK}$ of the dilution refrigerator. We extract a charging energy $\Delta E_C=3.5\,\textrm{meV}$ which is comparable to the values found in other devices of similar size \cite{Stampfer08}.



\begin{figure}
  \centering
  \includegraphics[width= 12 cm]{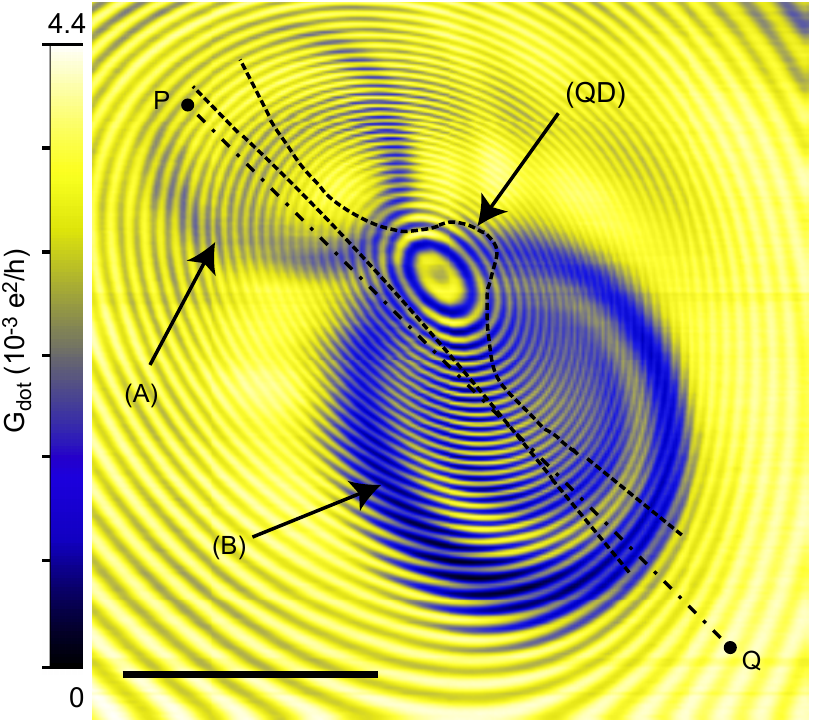}
  \caption{Scanning-gate image in the hole regime, $V_{\textrm{BG}}=12\,\textrm{V}$. The tip voltage was $V_{\textrm{tip}}=2\,\textrm{V}$, left gate voltage $V_{\textrm{LG}}=0.15\,\textrm{V}$,  and the scan frame has a size of $1.4\times 1.4\,\mu\textrm{m}^2$. A symmetric bias of $V_{\textrm{bias}}=300\,\mu\textrm{V}$ was applied across source and drain and the tip was scanned at a constant height of $\Delta z \approx 120\,\textrm{nm}$ above the sample. Coulomb resonances of the quantum dot show up as concentric rings denoted by arrow (QD). The center of the Coulomb resonances are offset from the topographic center of the dot by ca. $240\,\textrm{nm}$. Such a behavior, known from previous scanning-gate experiments, is understood and of minor importance here \cite{Gildemeister07b}. The outline of the quantum dot and its connection to source and drain via the two constrictions, depicted with dashed, black lines, is corrected for the offset, assuming that the Coulomb resonances are centered in the quantum dot (see also supporting information \cite{supp_info}). Most striking, however, is the appearance of two more sets of concentric rings which are highlighted by arrows (A) and (B) and which are centered around points in the constrictions. The black, dash-dotted line between points P and Q denotes the line along which the linescan of Fig.~3 was taken. The scale bar denotes $500\,\textrm{nm}$.\label{fig2}}
\end{figure}

We performed scanning-gate measurements of the quantum dot in the hole regime at $V_{\textrm{BG}}=12\,\textrm{V}$ (see supporting information \cite{supp_info} and Ref.~\citenum{Gildemeister07} for further information on our scanning-gate setup). To this end, the conductance $G_{\textrm{dot}}$ of the quantum dot is recorded as the voltage-biased tip is scanned at constant height above the structure \cite{Topinka01, Pioda04, Gildemeister07b}. A representative result is shown in Fig.~2; further images in the same regime are presented in the supporting information \cite{supp_info}. The scan frame has an area of $1.4\times 1.4\,\mu\textrm{m}^2$ and the outline of the quantum dot, as obtained from topographical images (see Fig.~1(b)), is shown with dashed, black lines. We observe three sets of concentric rings which are marked by arrows labelled (QD), (A), and (B). The set (QD) is caused by Coulomb resonances of the quantum dot as verified by the presence of Coulomb-blockade diamonds when sweeping the tip and bias voltages (not shown here) and we refer to them as \emph{Coulomb rings}. The conductance $G_{\textrm{dot}}$ does not drop to zero between two Coulomb rings because the measurements were done at the edge of the transport gap in backgate voltage where the coupling of dot states to source and drain is rather strong.

Most strikingly, we observe two additional sets of rings (A) and (B). In the following, we will refer to them as \emph{resonances} A and B, respectively. In all scanning-gate images taken on this sample, resonances A and B are manifest as amplitude-modulations of the Coulomb resonances of the quantum dot. They are centered around points in the constrictions connecting the quantum dot to source and drain. Their presence allows to locate regions of localized charge carriers in the constrictions. This is a central result of this paper. The interpretation of rings A and B as being due to \emph{localized states} will be corroborated below. Only one or at maximum two localized states are observed in each constriction.


\begin{figure}
  \centering
  \includegraphics[width= 12 cm]{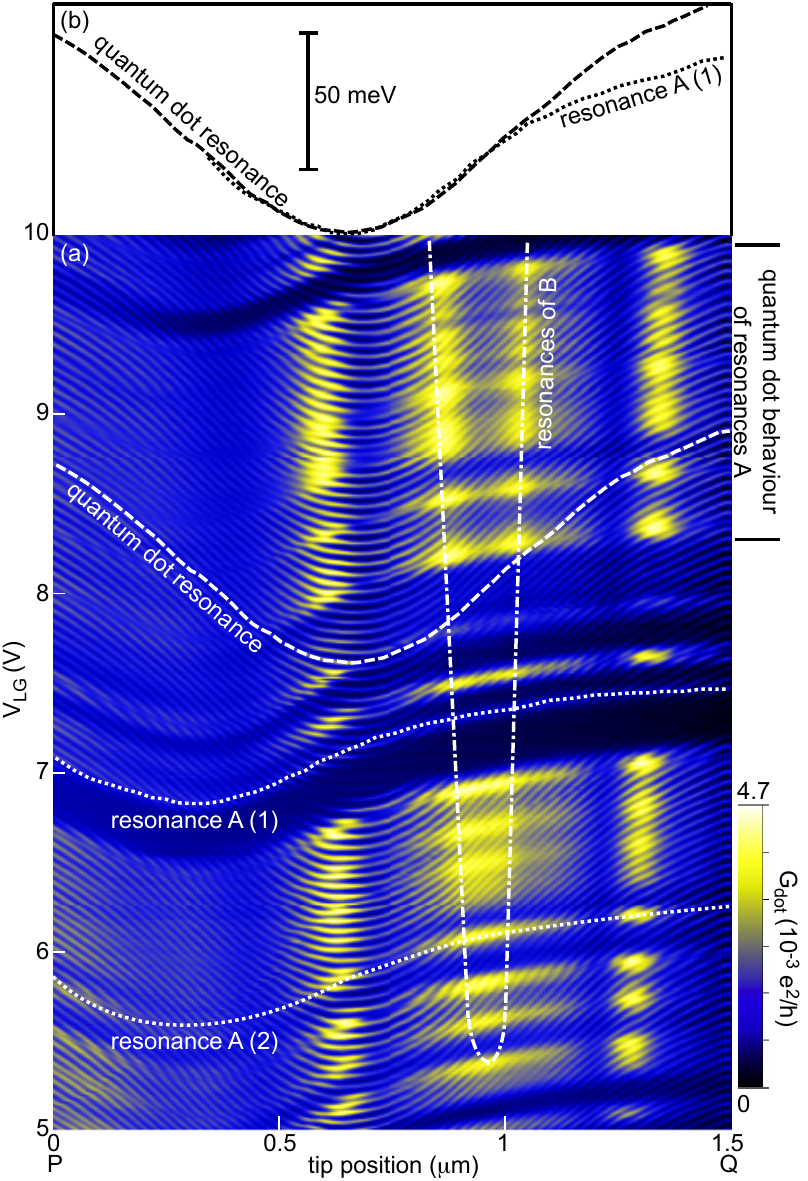}
  \caption{(a) Linescans along the dash-dotted line between points P and Q depicted in Fig.~2 while the left gate was stepped from $5\,\textrm{V}$ to $10\,\textrm{V}$. The other parameters are the same as in Fig.~2. We can clearly distinguish between features of resonances A and B and the quantum dot. Two resonances of A are highlighted with white, dotted lines labelled (1) and (2), a quantum dot resonance is highlighted with a white, dashed line, and a resonance of B ist denoted with a white, dash-dotted line. The upper, broad resonance (1) of A does not show any charging effect, whereas the lower, sharp one (2) is accompanied with avoided crossings of the Coulomb resonances. (b) Shifting resonance (1) of A along the x-axis and scaling it by 1.67 lead to the determination of relative lever arms between quantum dot and resonance A as explained in the text.\label{fig3}}
\end{figure} 

In electronic transport, resonances are in general localized in space and sharp in energy. In order to confirm this for resonances A and B, we took linescans between points P and Q in Fig.~2 and changed the energy of the localized states by stepping the left side gate voltage $V_{\textrm{LG}}$ from $5\,\textrm{V}$ up to $10\,\textrm{V}$. The result is shown in Fig.~3(a). We can identify quantum dot resonances (dashed, white line and resonances parallel to it), resonances A (dotted, white lines and resonances parallel to them), and resonances B (dash-dotted line and resonance parallel to it).

The tip-induced potential at any fixed location in the graphene plane is changed when moving the tip from P to Q. We stay on a particular resonance by compensating for this change at the location of the resonance with $V_{\textrm{LG}}$. This leads to the characteristic slopes of resonances A and B and the quantum dot resonances. The left side gate is closest to the center of resonance A; resonance B is furthest away. Therefore resonances A are strongly tuned by the left gate, whereas resonances B are only slightly affected. Quantum dot resonances are in between. Below we will use this effect to deduce the lever arm ratio $\alpha_{\textrm{LG}}^{\textrm{loc}}/\alpha_{\textrm{LG}}^{\textrm{dot}}$ of the lever arms of the left gate on the dot, $\alpha_{\textrm{LG}}^{\textrm{dot}}$, and on resonance A, $\alpha_{\textrm{LG}}^{\textrm{loc}}$.

Horizontal cuts in Fig.~3(a), i.~e. cuts for fixed $V_{\textrm{LG}}$, show that all resonances eventually shrink to a single point in space. This allows us to identify them with states localized in space at this point. Vertical cuts, on the other hand, show that they are reasonably sharp in energy. However, we notice a strong variation in the width of the resonances; e.~g. resonance A(2) is much sharper than resonance A(1). A closer inspection of the sharper resonances of A reveals that they are accompanied with avoided crossings of the quantum dot Coulomb resonances.


\begin{figure}
  \centering
  \includegraphics[width= 12 cm]{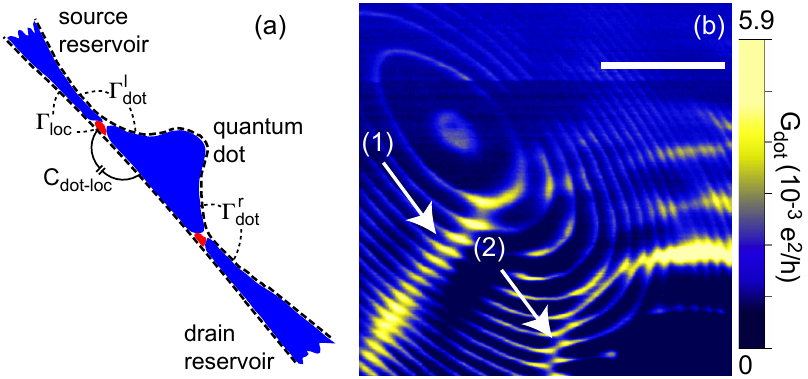}
  \caption{(a) Model for explaining the essential features induced by localized states in the constrictions. For simplicity, we consider just one localized state (upper red puddle) which is coupled to the source lead via the tunnel coupling $\Gamma_{\textrm{loc}}$. $C_{\textrm{dot-loc}}$ denotes the capacitive coupling between the localized state and the quantum dot. (b) Scanning-gate measurements at $T\approx 90\,\textrm{mK}$ measured at the mixing chamber. The image was taken at $V_{\textrm{BG}}=12\,\textrm{V}$, $V_{\textrm{bias}}=35\,\mu\textrm{V}$, $V_{\textrm{tip}}=-100\,\textrm{mV}$, and a tip height of $\Delta z=40-45\,\textrm{nm}$; the scan frame covers an area of $0.3\times 0.3\,\mu\textrm{m}^2$. Compared to Fig.~2, Coulomb resonances are much sharper due to the lower temperature. Although this measurement and Fig.~2 and Fig.~3 were taken at the same backgate voltages, a direkt comparison of them is difficult because of several charge rearrangements in between. The crossings of Coulomb resonances and resonances of B lead to a modulation of $G_{\textrm{dot}}$ and no avoided crossing of Coulomb resonance for arrow (1) and to avoided crossings for arrow (2). The scale bar denotes $100\,\textrm{nm}$.\label{fig4}}
\end{figure}

We propose the model shown in Fig.~4(a) which is capable of capturing the essential details of our observations. It consists of a quantum dot coupled to source and drain via two tunnel barriers with tunnel coupling $\Gamma_{\textrm{dot}}^{\textrm{l,r}}$. We introduce an additional localized state located in the constriction and coupled to the lead via a tunnel barrier with coupling $\Gamma_{\textrm{loc}}$. The localized state interacts with the quantum dot states by tunneling through the barrier and by mutual capacitive coupling via $C_{\textrm{dot-loc}}$.

Fig.~4(b) shows a scanning-gate image taken at the temperature $T\approx 90\,\textrm{mK}$. Coulomb resonances are now much sharper than in Fig.~2 and Fig.~3(a). Resonances B lead to a modulation of the dot conductance $G_{\textrm{dot}}$ as highlighted by arrows (1) and (2). The capacitive coupling between resonance B and the quantum dot leads to the avoided crossings pointed at by arrow (2). They are more easily identified compared to Fig.~3(a) because of the lower temperature (see supporting information for a further scanning-gate image in this regime \cite{supp_info}).


Whenever such an avoided crossing occurs, resonance A or B is charged with an additional charge carrier. Thus resonances originate from localized states. The capacitive coupling $C_{\textrm{dot-loc}}$ shifts the chemical potential in the dot when the localized state is charged. Consequently, the Coulomb ring in the scanning-gate image is shifted as well. Charging effects are not observed for all crossings of quantum dot resonances with resonances A or B. Avoided crossings of Coulomb resonances occur only for narrow resonances A and B; the broader ones do not show the signature of charge quantization as inspection of Fig.~3 and Fig.~4 show. The tunnel coupling strength $\Gamma_{\textrm{loc}}$ must therefore depend strongly on the Fermi energy. Then charging of a localized state with discrete charges occurs only if $\Gamma_{\textrm{loc}}$ is below a certain threshold such that the condition $G_{\textrm{loc}}\left(\Gamma_{\textrm{loc}}\right)<e^2/h$ for the conductance of the localized state is fulfilled. The width of the resonance is also determined by the tunnel coupling if $\Gamma_{\textrm{loc}} > 4k_BT$.

Whenever a localized state shows quantum dot-like behavior, we can infer its size from its charging energy. In order to do so, we pick just those resonances of A in Fig.~3  which show avoided crossings with quantum dot resonances. We identify six resonances of this type in the regime denoted by ``quantum dot behavior of resonances A''. They have a spacing of $\Delta V_{\textrm{LG}}^{\textrm{loc}}\approx 270\,\textrm{mV} - 370\,\textrm{mV}$ in left-gate voltage. In order to convert this voltage scale to an energy, we need to know the lever arm $\alpha_{\textrm{LG}}^{\textrm{loc}}$ of the left gate onto the localized state. Coulomb resonances of the quantum dot are separated on average by $\Delta V_{\textrm{LG}}^{\textrm{dot}}=53\,\textrm{mV}$ in left-gate voltage. Combined with the charging energy $\Delta E_C = 3.5\,\textrm{meV}$, this yields a lever arm of $\alpha_{\textrm{LG}}^{\textrm{dot}}=0.066$. The two dotted lines in Fig.~3(a) denote how resonances of A evolve as a function of position. If we scale these lines by a factor of 1.67 in $V_{\textrm{LG}}$-direction and shift them along the x-axis (the ``position'') such that their minima coincide with a minimum of a Coulomb resonance, they nicely fit onto each other over a range of almost $1\,\mu\textrm{m}$ as shown in Fig.~3(b). (For tip positions $> 1.1\,\mu\textrm{m}$, screening effects lead to a different progression of the quantum dot resonance and the resonances of localized state A.) This gives us the possibility to estimate the desired lever arm $\alpha_{\textrm{LG}}^{\textrm{loc}}$. It is given by $\alpha_{\textrm{LG}}^{\textrm{loc}}=\alpha_{\textrm{LG}}^{\textrm{dot}}\cdot 1.67 = 0.110$. Consequently the charging energy $\Delta E_C^{\textrm{loc}}=\alpha_{\textrm{LG}}^{\textrm{loc}}\cdot e \Delta V_{\textrm{LG}}^{\textrm{loc}}$ of the localized state A is between $30\,\textrm{meV}$ and $41\,\textrm{meV}$. This corresponds to a radius of the localized state of $r_{\textrm{loc}}=10\,\textrm{nm} - 13\,\textrm{nm}$ assuming $\Delta E_C^{\textrm{loc}}/\Delta E_C=r/r_{\textrm{loc}}$ with $\Delta E_C = 3.5\,\textrm{meV}$ and lithographic size of the dot $r=110\,\textrm{nm}$. Previous experiments obtained similar sizes by means of conventional transport experiments for graphene nanoribbons \cite{Todd08, Molitor09, Han09}. However our method allows to determine relative lever arms of localized states in more complex structures such as the presented quantum dot device.

Stampfer \emph{et al.} report on a variation of relative lever arms of localized states in nanoribbons by up to 30\% \cite{Stampfer09}. This is explained with a number of localized states spread along the nanoribbon. Based on the geometry of the device used in \citenum{Stampfer09}, a rough estimate yields a spread of over 300 nm. With our scanning-gate microscope, a spread of more than 100 nm should easily be detectable. The fact that we observe one localized state per constriction is most likely due to the short length of the constrictions.

\begin{acknowledgement}
  Financial support by ETH Z\"urich and the Swiss National Science Foundation is gratefully acknowledged. Some images have been prepared using the WSxM-software \cite{Horcas07}.
\end{acknowledgement}

\begin{suppinfo}
  The supporting information includes description on the sample preparation and of our scanning-gate setup. Additionally more scanning-gate images are shown to corroborate some statements of this letter.
\end{suppinfo}



\begin{tocentry}
  \includegraphics{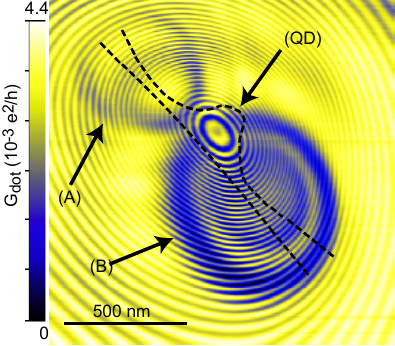}
\end{tocentry}

\end{document}





The graphene sample has been fabricated by mechanical exfoliation of natural graphite and subsequent deposition on a highly doped silicon wafer covered by 285 nm of silicon dioxide. Raman spectroscopy confirmed that the flake consisted of a single atomic layer. A final e-beam step is used to pattern Ti/Au electrodes. A detailed description of the fabrication process is given in Ref.~\citenum{Stampfer08}. Fig.~1(a) in the article shows an atomic force micrograph of the quantum dot (QD) and nearby in-plane gates (left and right side gate, LG and RG, and plunger gate PG) under ambient conditions after etching and removing the protective resist layer. The image was taken under ambient conditions. The quantum dot has a lithographic radius of $r\approx 110\,\textrm{nm}$; the two constrictions have a width of 30 to 40 nm.


We equipped an Oxford Instruments Kelvinox 100 dilution refrigerator with a home-built atomic force microscope (AFM) \cite{Gildemeister07}. Quartz tuning forks with etched PtIr-tips are used as force sensors. The completed device shown in Fig.~1(b) in the main article was imaged in-situ with this microscope at $T\approx 2.6\,\textrm{K}$. The topographic image shows clear edges at the gold contacts and even the graphene nanoribbon edges can be seen as faint lines. This indicates the use of a sharp tip. In order to obtain a scanning-gate image, the AFM-feedback is turned off and a constant voltage $V_{\textrm{tip}}$ is applied to the tip. The current $I_{\textrm{dot}}$ through the dot is then recorded as a function of tip position while scanning the tip at constant height above the sample \cite{Topinka01, Pioda04, Gildemeister07b}. Coulomb resonances of the quantum dot occur whenever the tip-induced potential shifts an energy level of the quantum dot into resonance with the electrochemical potential of source and drain. Hence, the rings can be regarded as contour lines of constant electrochemical potential in the dot \cite{Gildemeister07b}. The energy difference of neighbouring contour lines is $\Delta E_C$, the charging energy of the dot. The contour line pattern reflects the tip-induced potential as sensed by the quantum dot states. 

The excellent electronic quality of the tip can be deduced from the Coulomb rings. In Fig.~2 in the main article, we can distinguish more than 30 Coulomb rings. Previous experiments showed features which had to be attributed to double- or multi-tip behavior on smaller length scales than in the images shown here \cite{Pioda04, Gildemeister07b, Gildemeister07c}. We can resolve neighbouring Coulomb rings which are separated by less than 20 nm.


\begin{figure}[t]
  \includegraphics[width=.5 \linewidth]{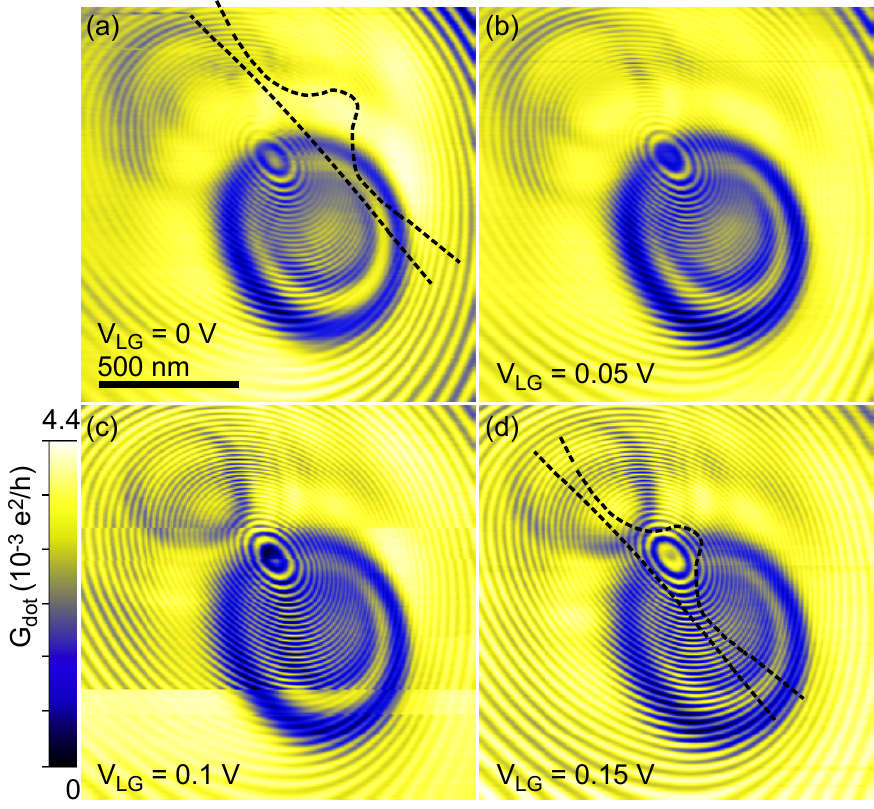}
  \caption{(Figure S1) Scanning-gate images for the same parameters as in Fig.~2 of the main article for different voltages of the left side gate. In the main article, figure (d) from above is presented. (a) The outline of the quantum dot and its position relative to the scanning-gate image is depicted as obtained from topographic scans. (d) The outline of the quantum dot has been shifted by 240 nm such that the Coulomb resonances are centered in the topographic center of the dot. \label{fig_S1}}
\end{figure}

Fig.~S1 shows scanning-gate images for four different side gate voltages. These images were taken over several hours and prove the stability of the sample. In (a), the outline of the quantum dot is depicted as black, dashed lines at the position according to the AFM scan in Fig.~1(b) of the main paper. The position of the outline is shifted by 240 nm in (d) such that the Coulomb resonances are centered in the quantum dot. As pointed out in the main article, a difference between the electronic and the topographic tip is known from previous experiments.

This series of four scanning-gate images also show that we can tune the resonances of the localized states in the same way as the quantum dot resonances by changing gate or tip voltages.

We present a further scanning-gate image in Fig.~S2 taken in the same regime as Fig.~4(b) in the main article but with $V_{\textrm{PG}}=6\,\textrm{V}$. The arrow highlights an avoided crossing of Coulomb resonances. The Coulomb resonances show a wobbling behavior which we did not observe before. The origin is not completely clear to us but it might be due to quantum mechanical confinement effects.

\begin{figure}
  \includegraphics[width=.5 \linewidth]{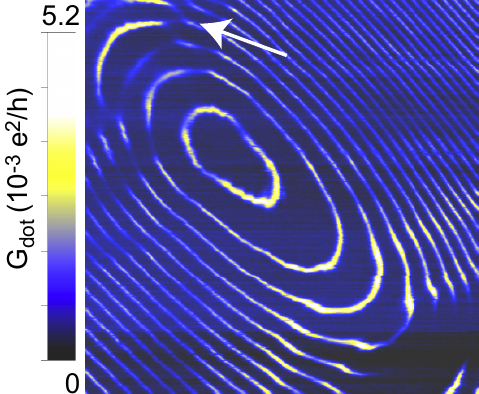}
  \caption{(Figure S2) Scanning-gate image for the same parameters in in Fig.~4 of the main article and $V_{\textrm{PG}}=6\,\textrm{V}$. The arrow points to an avoided crossing of the Coulomb resonances. Additionally Coulomb resonances show a wobbling behavior unobserved before. \label{fig_S2}}
\end{figure}